\documentclass{aa} 
\usepackage[dvipsnames]{xcolor}
\usepackage{txfonts}

\definecolor{tiffany}{RGB}{79, 166, 158}

\newcommand{\higpus}{\texttt{HiGPUs~}}
\newcommand{\arwv}{\texttt{ARWV~}}
\newcommand{\nemo}{\texttt{NEMO~}}

\usepackage{natbib}
\bibpunct{(}{)}{;}{a}{}{,}

\usepackage{graphicx}
\usepackage{amsmath}
\usepackage{grffile}

\DeclareMathOperator{\erf}{erf}

\begin{document}

\title{Future merger of the Milky Way with the Andromeda galaxy and the fate of their supermassive black holes}

\author{Riccardo Schiavi
\inst{1}
\and
Roberto Capuzzo-Dolcetta \inst{1}
\and
Manuel Arca-Sedda \inst{2}
\and
Mario Spera \inst{3, 4, 5}}

\institute{Dipartimento di Fisica, Universit\`a degli Studi di Roma ``La Sapienza'', P.le Aldo Moro, 2, I-00185 Rome, Italy \\
\email{riccardo.schiavi@uniroma1.it}
\and
Astronomisches Rechen Intitut, Zentrum f\"ur Astronomie der Universit\"at Heidelberg, M\"onchhofstrasse 12-14, D-69120 Heidelberg, Germany 
\and
Universit\`a di Padova, Dipartimento di Fisica e Astronomia, I-35131 Padova, Italy
\and
INFN, Sezione di Padova, I-35131 Padova, Italy
\and
Center for Interdisciplinary Exploration \& Research in Astrophysics (CIERA), Northwestern University, Evanston, IL 60208, USA
}

\date{}

\abstract{Our Galaxy and the nearby Andromeda galaxy (M31) are the most massive members of the Local Group, and they seem to be a bound pair, despite the uncertainties on the relative motion of the two galaxies. A number of studies have shown that the two galaxies will likely undergo a close approach in the next 4$-$5 Gyr. We used direct $N$-body simulations to model this interaction to shed light on the future of the Milky Way - Andromeda system and for the first time explore the fate of the two supermassive black holes (SMBHs) that are located at their centers. We investigated how the uncertainties on the relative motion of the two galaxies, linked with the initial velocities and the density of the diffuse environment in which they move, affect the estimate of the time they need to merge and form ``Milkomeda''. After the galaxy merger, we follow the evolution of their two SMBHs up to their close pairing and fusion. Upon the fiducial set of parameters, we find that Milky Way and Andromeda will have their closest approach in the next 4.3 Gyr and merge over a span of 10 Gyr. Although the time of the first encounter is consistent with other predictions, we find that the merger occurs later than previously estimated. We also show that the two SMBHs will spiral in the inner region of Milkomeda and coalesce in less than 16.6 Myr after the merger of the two galaxies. Finally, we evaluate the gravitational-wave emission caused by the inspiral of the SMBHs, and we discuss the detectability of similar SMBH mergers in the nearby Universe ($z\leq 2$) through next-generation gravitational-wave detectors.}
\authorrunning{Schiavi et al.}
\titlerunning{Future merger of the MW and M31 and the fate of their SMBHs}
\maketitle

\section{Introduction}

The Milky Way (MW) and the Andromeda galaxy (M31) are the two main members of the Local Group, which contains more than 80 galaxies and has a total mass of roughly $3-5\times10^{12}\,\mathrm{M_{\odot}}$ \citep{Gonzalez14,VdM12a}. The future evolution of the Local Group is essentially driven by the dynamics of our Galaxy and M31, and it can be considered a promising study object to investigate the processes of galaxy formation and evolution. Although the physical and dynamical properties of the MW-M31 system are rather uncertain, it is likely that the Local Group is gravitationally bound and decoupled from the general cosmic expansion, and also that the two galaxies will not escape the collision and the final merger. However, the time at which this merger will occur is still a matter of debate. The main purpose of this work and of our previous studies \citep{schiavi18,schiavi19} is to shed light on this topic. \\
According to some previous simulations \citep{dubinski96,Coxloeb08,VdM12b}, the first close approach will likely occur in $<4$ Gyr, even though the two galaxies have different initial conditions in all the cited works. Using a more recent estimation of the proper motion of M31, \citet{VdM19} have obtained a time for the first approach equal to $4.5$ Gyr.  Almost the same result can also be obtained when the so-called ``timing argument'' is employed, which was introduced in the pioneering work by \citet{KW59}. In the timing argument, MW and M31 are considered as point masses on a radial orbit: they started their motion at the Big Bang, and after decoupling from the Hubble flow, began to approach one another. Even though the timing argument allowed obtaining an estimate of the total mass of the Local Group that is compatible with the estimate obtained with other methods \citep[e.g.,][]{klypin02,wdubinsky05}, it is unable to take the complexity of the dynamics of the galaxy interaction into account. The time needed for the completion of the whole merger process is highly sensitive not only to the masses of the two galaxies, but also to their proper motion and to the density of the intergalactic medium (IGM) in which they move. 
All the estimates of the mass of the two galaxies are affected by a rather high level of uncertainty. This is due mainly to the presence of extended dark matter halos. We have chosen to adopt the values estimated in \citet{klypin02}, defined as the virial masses at radius $r_{200}$, where the galactic density is 200 times the critical density $\rho_{0}\approx1.0\times10^{-26}\,\mathrm{kg/m^3}$ (according to the measurements of the Hubble constant $H_0$ by \citet{huang20} and \citet{planck18}): $M_{MW}=1.0\times10^{12}\,\mathrm{M_{\odot}}$ and $M_{M31}=1.6\times10^{12}\,\mathrm{M_{\odot}}$. Another source of uncertainty is our poor knowledge of the actual size of the two galactic halos. The radial extent of the halo in equilibrium models of galaxies developed by \citet{Kuij_Dub95}  is in the range of $21-73$ disk scale lengths. The ratio of the dark halo virial radius and the galaxy effective radius fall in the same range in the studies by \citet{Jiang_2019}, \citet{Somerville17}, \citet{huang17}, and \citet{Kravtsov_2013}. However, there is evidence that some of the gaseous circumgalactic medium (CGM) extends to even larger distances \citep{shull14}. In other words, we do not know with sufficient precision where a galaxy actually ends, and in our specific case, whether the Milky Way and Andromeda are already partially overlapping or if they are still well separated. For this reason, as discussed in Section \ref{model}, we have decided to set the halo cutoff radii of the two galaxies at the respective tidal radii. Moreover, in Section \ref{gal_merge} we demonstrate that the time of the merger does not depend on galactic halos that are more extended than 80 disk scale lengths. Because it is evident that the edge of each galaxy gradually fades in the IGM, we cannot ignore the effect of this diffuse medium in studying the interaction. The density of the IGM is known to be $\text{four to six}$ times the critical density $\rho_{0}$ \citep{takodoro71,Coxloeb08}, but by performing several simulations, we obtained that even a small variation in this parameter could affect the merger time substantially. \\ 
Our knowledge of the proper motion of M31 relative to us is mainly obtained through redshift measurement. This gained us an accurate estimate of the only radial component of the relative velocity vector of M31: $V_{r}\approx120$ km/s \citep{binney87}. The tangential component has been inferred by studying the motion of the satellite galaxies of M31 \citep{loeb05,VdM08} or by the Hubble Space Telescope (HST) and GAIA observations of sources behind M31 \citep{VdM12b,VdM14,VdM16}. These estimates span from a minimum of $V_{t}\approx17$ km/s \citep{VdM12a} to a maximum of $V_{t}\approx164$ km/s \citep{salomon16}. The most recent estimate is that by \citet{VdM19}, who have used GAIA DR2 to obtain $V_{t}=57_{-31}^{+35}$ km/s. We referred to this ultimate measurement to fix the orientation of the relative velocity vector and its radial component ($V_{r}=-115.7$ km/s). We discuss the relation between the initial tangential velocity and the time of the merger in Section \ref{gal_merge}. \\
During the interaction at large scales, we are interested in following the motion of the two SMBHs in the centers of the two galaxies. It is well known that a compact object of mass $M_{MW}^{\bullet}=4.31\times10^6\,\mathrm{M_{\odot}}$ \citep{gillessen09}, called SgrA*, is placed at the center of the Milky Way. Even though the nucleus of M31 seems to have a double or triple structure, there is a high probability that it might host an SMBH of mass $M_{M31}^{\bullet}=1.4\times10^8\,\mathrm{M_{\odot}}$ \citep{bender05}.
After the merger of the two host galaxies, their SMBHs are expected to form a binary that will shrink over time through gravitational encounters with field stars. We first explore the future evolution of some of the nearest SMBHs and their eventual coalescence in the nucleus of the galactic merger remnant. In Section \ref{smbh_merge} we discuss the time required for the SMBHs to merge and the amount of energy radiated through gravitational waves (GWs). \\
One of the most effective ways to model galaxy interactions is the integration of the $N$-body problem. While tree codes can simulate a large number of collisionless particles in galaxy merger simulations, a collisional direct summation $N$-body code with fewer particles is required in this study to follow the SMBH dynamics. Direct $N$-body codes are highly reliable but computationally expensive: this clearly places a limit on the number of particles involved in the simulations, and prevents us from resolving large and small scales at the same time with good accuracy. For this reason, as we discuss in Section \ref{methods}, we chose to split the whole study into two parts: in the first, we simulate the galaxy interaction at large scales, and in the second, we focus on the analysis of the orbital decay of the SMBH binary.

\section{Galactic model and initial conditions} \label{model}
Our galaxies were modeled by combining three different components: an exponential disk, a spherical bulge, and a halo, the latter two with a Hernquist \citep{hernquist90} density profile. We combined these three components with the command {\fontfamily{qcr}\selectfont magalie} in the \nemo code \citep{teuben95}, which guarantees the stability of the whole system.\\ 
The two galaxies have the structure presented in \citet{klypin02} and in \citet{wdubinsky05}, that is, nearly the same as  was used by \citet{Coxloeb08}. The main structure parameters are summarized in Table \ref{tab1}.

 \begin{table}[h]
 \centering
  \begin{tabular}{l c c}
  \hline\hline 
  & {\bf Milky Way} & {\bf Andromeda} \\ \hline
{\bf Scale radius of the disk ($\mathrm{kpc}$)} & $3.5$ & $5.7$ \\ \hline
{\bf Core radius of the bulge} & $0.2$ & $0.2$ \\ \hline
{\bf Core radius of the halo} & $3.0$ & $3.0$ \\ \hline
{\bf Cutoff radius of the halo} & $98.3$ & $76.5$ \\ \hline
{\bf Mass of the disk ($\mathrm{M_{\odot}}$)} & $4.0\times10^{10}$ & $7.0\times10^{10}$ \\ \hline
{\bf Mass of the bulge} & $0.2$ & $0.3$ \\ \hline
{\bf Mass of the halo} & $23.8$ & $21.6$ \\ \hline
{\bf Total mass ($\mathrm{M_{\odot}}$)} & $1.0\times10^{12}$ & $1.6\times10^{12}$ \\ 
\hline\hline
\ \\
  \end{tabular}
 \caption{Values of characteristic parameters used in our simulations. Where it is not specified, the lengths are in units of the scale radius of the disk $R_d$ and masses are in units of the mass of the disk $M_d$.}
 \label{tab1}
 \end{table}
 
\begin{figure}[h]
\centering
\includegraphics[width=3.5in]{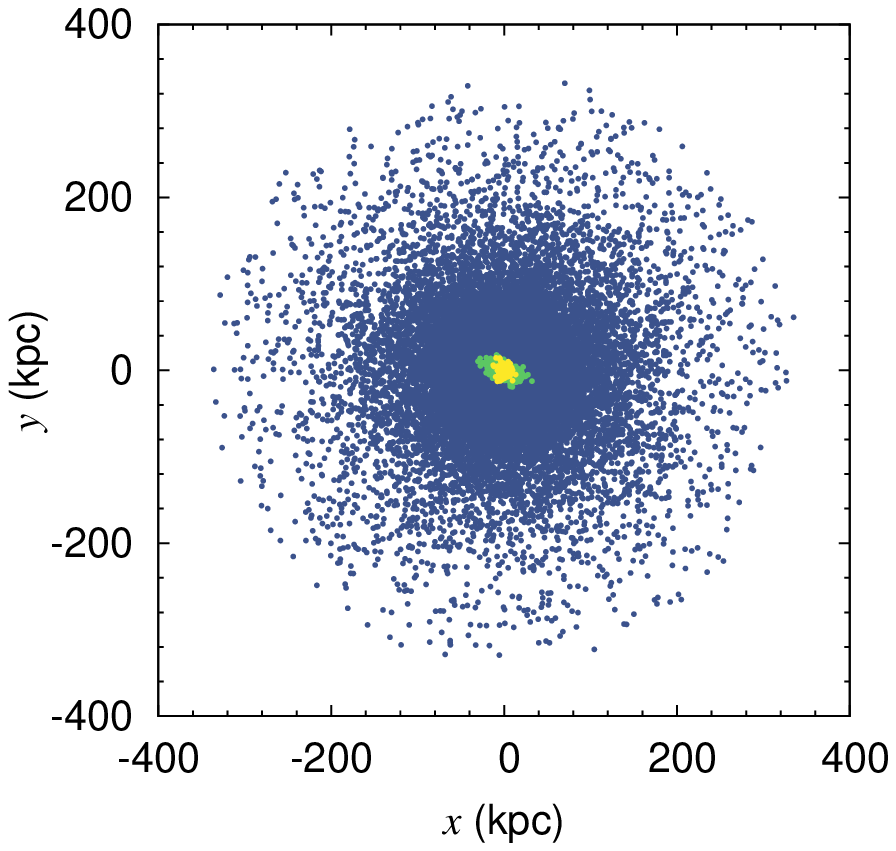} 
\includegraphics[width=3.5in]{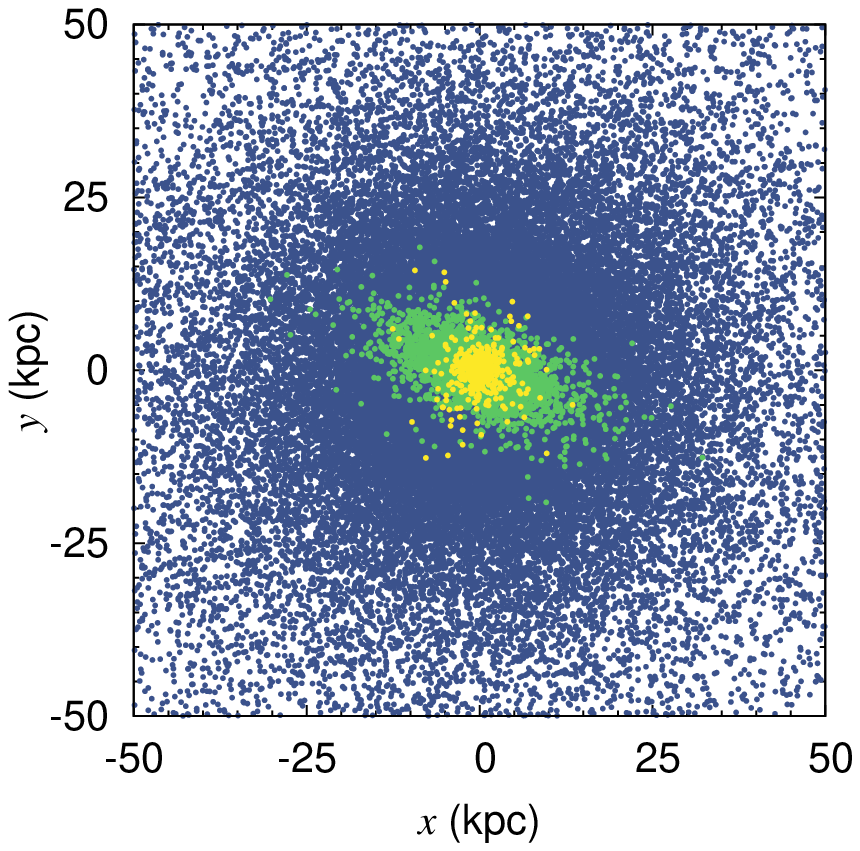} 
\caption{Our model of the Milky Way. The three  components (disk, bulge, and halo) are shown in different colors. The lower panel is a zoom into the innermost region.
}
\label{fig1}
\end{figure}

\begin{figure}[h]
\centering
\includegraphics[width=3.5in]{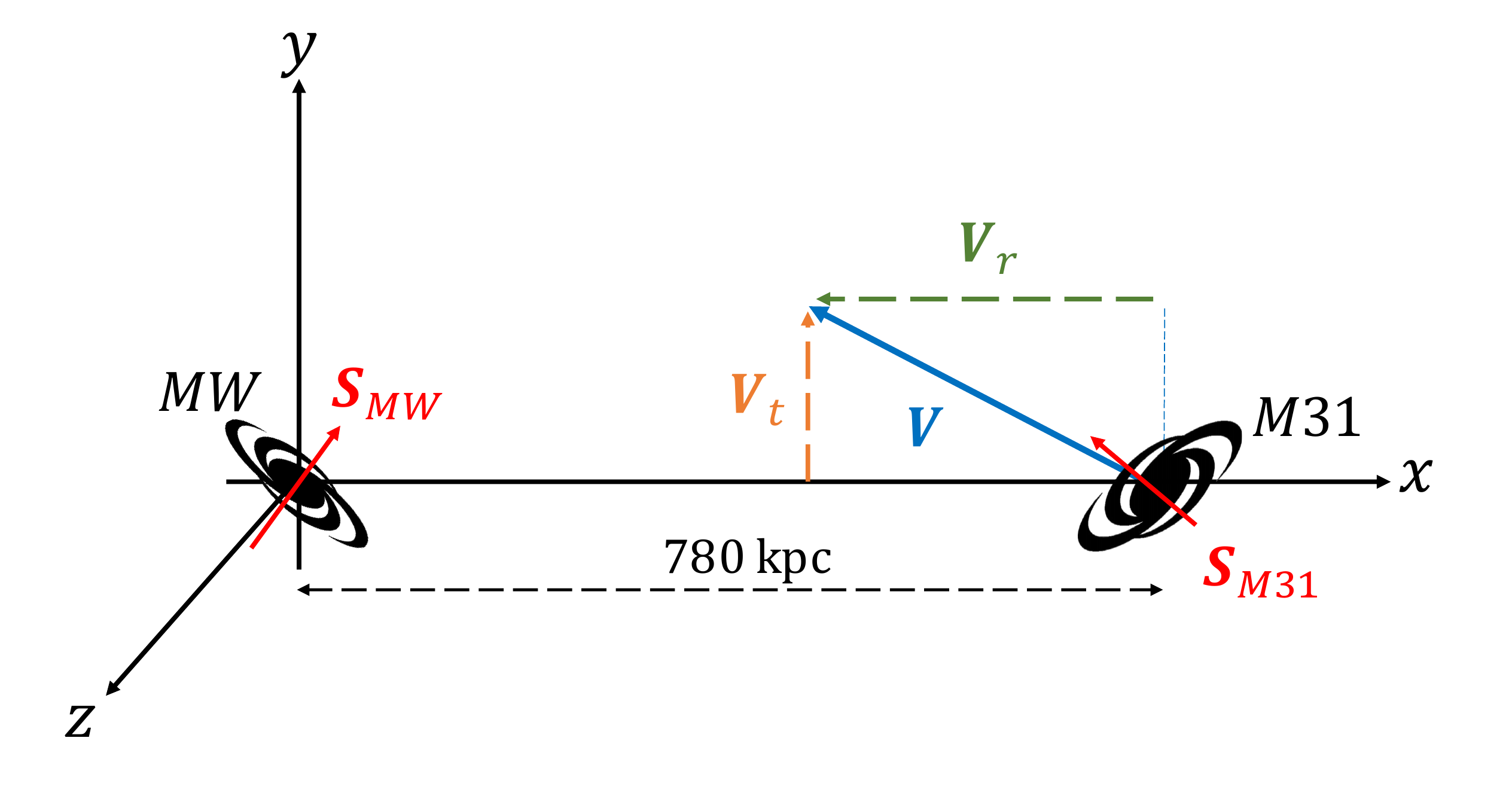}
\caption{Diagram of the initial configuration of the two galaxies in the chosen reference frame.}
\label{fig2}
\end{figure}

The cutoff radii of the halos were chosen as the tidal radii of the two galaxies, computed in the point-mass approximation:
\begin{equation}
\centering
    \frac{M_{MW}}{M_{M31}}=\left(\frac{r_{t,MW}}{R-r_{t,MW}}\right)^2\quad\text{and}\quad
    r_{t,M31}=R-r_{t,MW},
\end{equation}
where $r_{t,MW}$ and $r_{t,M31}$ are the two tidal radii and $R$ is the current separation between the two galaxies.
A snapshot of our galactic model is shown in Fig.\,\ref{fig1}.\\
We placed an SMBH in the center of each galaxy, which in the first part of our study was modeled as a particle with a mass of 0.001 times the mass of the whole galaxy. 
This ratio implies that the mass of our SMBHs is significantly higher than the observed masses, but this is not relevant for the dynamics of the two galaxies until merging because the two SMBHs are essentially passive guests of the hosting galaxies at this phase. We therefore made this choice because it was the best setting allowed by our numerical resolution. We used a number of particles not greater than $N=2.6\times10^5$ , and this constrains the mass of the single particle. An SMBH mass of one thousandth of the mass of the galaxy is therefore a good compromise between the properties of the galaxies and the comparison with an ordinary particle. However, during the simulation of the collision at large scales, the two particles that represent the SMBHs only have the purpose of better identifying the two galactic centers and of observing their relative distance at the end of the merger process.\\
The Milky Way and Andromeda start to interact at the current distance of 780 kpc \citep{mcconn05,Ribas05}, and their spin vectors are oriented at $(0^{\circ}; -90^{\circ})$ and $(240^{\circ}; -30^{\circ})$ in Galactic coordinates, respectively \citep{dubinski96,raylbell89}. To better display the dynamics of the galaxy binary system, we chose a reference frame where the x-y plane coincides with the plane of the motion. The initial configuration of the two galaxies is shown in Fig.\,\ref{fig2}.

\section{Methods} 
\label{methods}
Owing to the computational complexity of the simulations, we decided to divide the study into two parts. The first examines the galaxy interaction at large scales and has the main purpose of determining the true time required to form Milkomeda and its final density profile. In this section we use a number of particles of $N=2.6\times10^5$ for the simulations with the fiducial set of parameters and $N=6.5\times10^4$ for all the others. The numerical integration of the N-body equations of motion was implemented with the \higpus code \citep{Capuzzo2013}. This program is based on a sixth-order Hermite integration scheme with block time steps and directly computes the mutual force between each pair of particles, exploiting a parallelization of CPUs and GPUs. Owing to the high performance of the \higpus code, we repeated the simulation several times, changing two parameters that were linked with the initial conditions and the external environment: the tangential component of the initial relative velocity $V_{t}$ , and the density of the IGM $\rho$. This allowed us to investigate the correlation between the time of the merger and the galactic dynamical properties, together with the effect of the dynamical friction exerted by the surrounding diffuse medium. \\
In the second part we further study the evolution of the SMBH binary that formed after the galaxy merger. Taking the simulation with the highest resolution and the fiducial set of parameters, we obtained the Milkomeda density profile and the velocity dispersion and modeled the galactic center as an analytic distribution of matter around the SMBH binary. To simulate the evolution of the binary, we used a modified version of the \arwv code \citep{mikkola99,mikkola08}, which integrates the equations of motion taking in account the effect of the dynamical friction exerted by a diffuse background during the first phases of the orbital evolution, the post-Newtonian (PN) terms when the binary shrinks enough to reach the GW emission regime, and the effect of the spins of the merging objects \citep{arcasedda19,cha19}.
To connect the new simulation to the previous one, the SMBHs orbit was reproduced with the same geometry as found at the galaxy merger, while the masses of the two objects were set to those known from observational evidences.
Through this method, we can study the orbital decay of our binary down to small spatial scales and infer the coalescence time, the evolution of the semimajor axis, the eccentricity, and the power emitted in the form of GW.

\section {Intergalactic medium and dynamical friction}
\label{igm}
The presence of the IGM affects the time of the galaxy interaction through the extraction of orbital energy and angular momentum. We used \higpus to simulate the dynamics of the galaxy collision in different environments. To take the effect of the IGM into account, we modified the \higpus code by adding a dynamical friction term in the equation of motion of each particle according to the Chandrasekhar formula \citep{binney87}, 

\begin{multline}
   \frac{d^2\textbf{r}_i}{dt^2}=\sum\limits_{j\neq i}^N \frac{Gm\left(\textbf{r}_j-\textbf{r}_i\right)}{\left(\epsilon^2+\left|\textbf{r}_j-\textbf{r}_i\right|^2\right)^{3/2}}\\
   -\frac{4\pi G^2\rho M\ln{\Lambda}}{V_c^3}\left[\erf{(X)}-\frac{2X}{\sqrt\pi}e^{-X^2}\right]\textbf{V}_c,
   \end{multline}

with $X=V_c/\sqrt{2}\sigma$, where $\sigma$ is the IGM velocity dispersion.\\
As usual for $N$-body codes, we introduced the softening parameter $\epsilon$ to avoid the divergence of the Newton term at small distances: it was fixed at $\epsilon=500$ pc for ordinary particles and $\epsilon=50$ pc for the two black holes. In the Chandrasekhar term we considered $\rho$ as the density of the IGM, and $M$ and $V_c$ as the mass and velocity of the galactic core that the particle belongs to. Unlike the classical Chandrasekhar formula, which describes the effect of the dynamical friction on each star, depending on the stellar mass and velocity, in our case, each particle, which has the same mass $m$, feels the same frictional force as all the others. This force changes with time during the galaxy interaction, but at any moment, is the same for every particle. This choice is suggested by the need of describing the collective effect of the friction on the motion of each galaxy as a whole.\\
In all our simulations we fixed the Coulomb logarithm at $\ln{\Lambda}=5$ and the velocity dispersion of the medium at $\sigma=86.2\,\mathrm{km/s}$, obtained from the equipartition of energy for a diffuse medium at a temperature of $T=3\times10^{5}$ K \citep{Coxloeb08}.\\
We compared the case with no IGM with three cases with different values of $\rho$: $1.0\times10^{-26}\,\mathrm{kg/m^3}$, the same as the critical density, $4.0\times10^{-26}\,\mathrm{kg/m^3}$, which is the value estimated by \citet{takodoro71}, and $1.0\times10^{-25}\,\mathrm{kg/m^3}$, about 10 times the critical density. As we show in Section \ref{gal_merge}, the time of the merger significantly changes for different IGM densities, especially for high initial velocities.

\section{Results}
\subsection{Galaxy merger} \label{gal_merge}
\begin{figure}[h]
\centering
\includegraphics[width=3.5in]{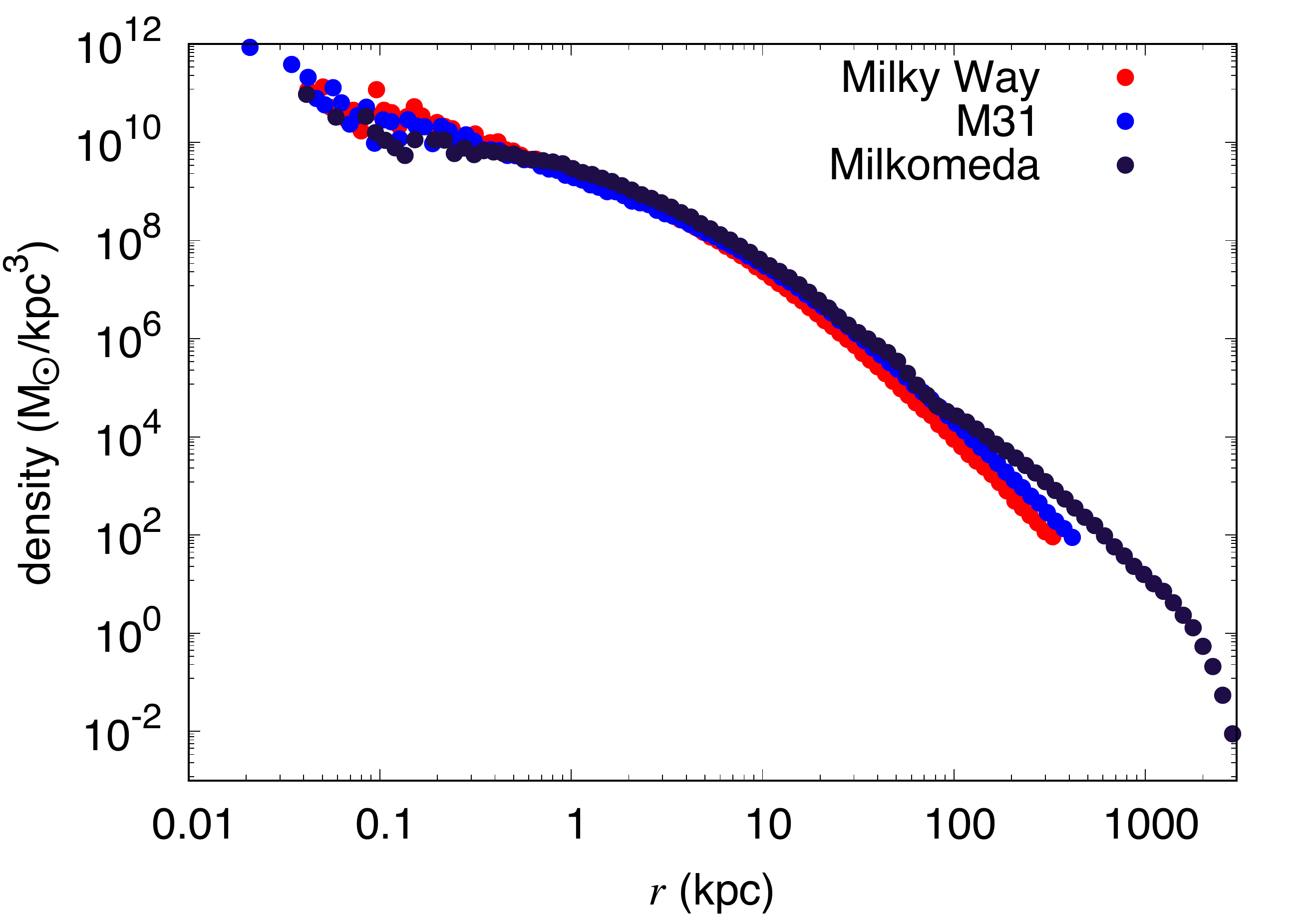}
\caption{Density profile of Milkomeda at time $t=10.2$ Gyr in the case with $V_t=57$ km/s and $\rho=4.0\times10^{-26}\,\mathrm{kg/m^3}$, and those of the Milky Way and M31 at $t=0$. }
\label{fig3}
\end{figure}

For all initial conditions, the merger remnant Milkomeda resembles a giant elliptical galaxy with a density profile similar to those of the original two galaxies, as is shown in Fig.\,\ref{fig3} for $V_t=57$ km/s and $\rho=4.0\times10^{-26}\,\mathrm{kg/m^3}$.

\begin{figure}[h]
\centering
\includegraphics[width=3.5in]{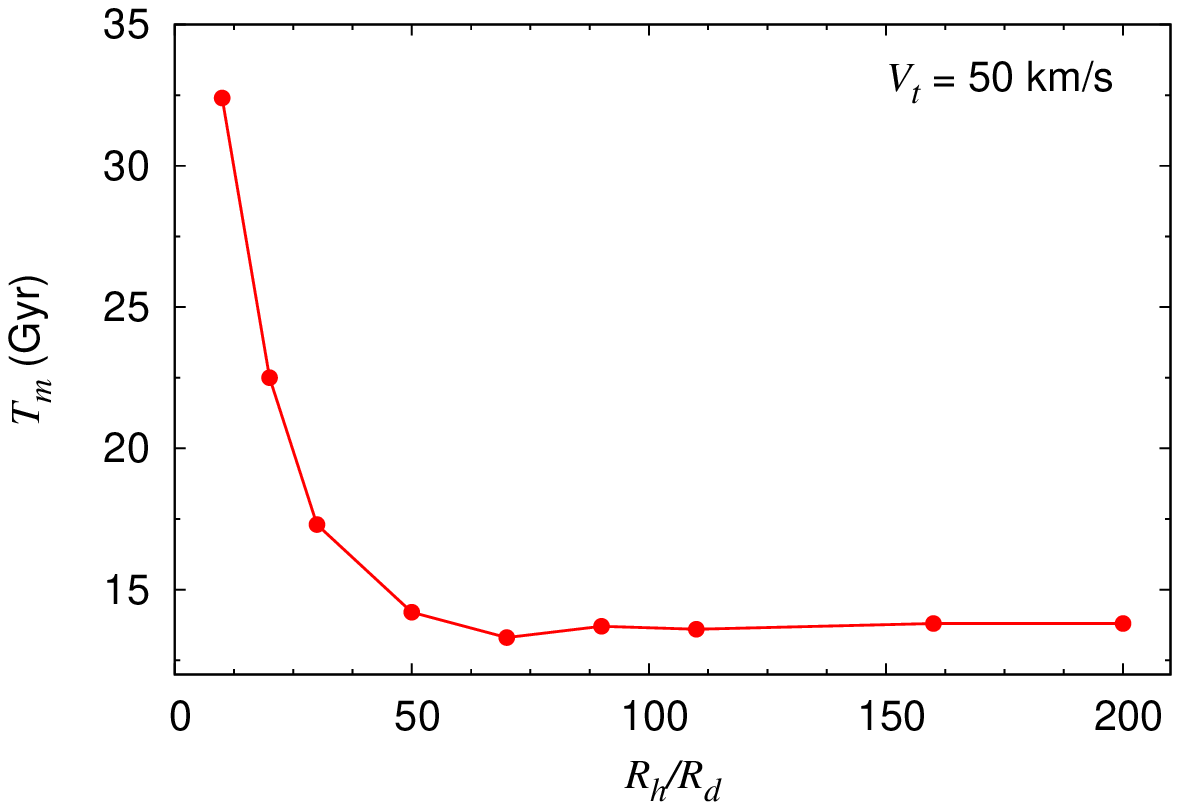}
\includegraphics[width=3.5in]{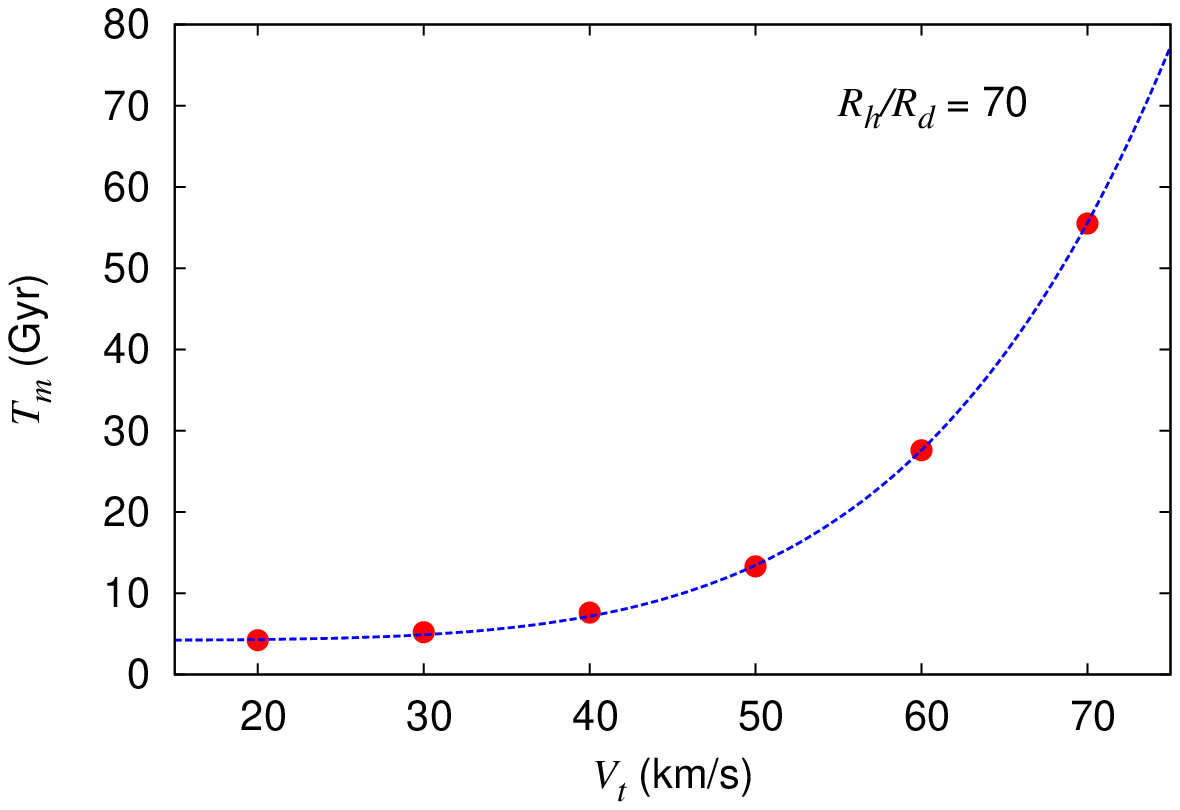}
\caption{Top: Time of the merger as a function of the cutoff radius of the halo $R_h$, given in units of the scale radius of the disk $R_d$ and assumed equal for the two galaxies. The initial tangential velocity here is fixed at $50$ km/s. Bottom: Correlation between the time of the merger and the initial tangential velocity in the case of no IGM. The cutoff radii of the halos here are fixed at $70\,R_d$.}
\label{time}
\end{figure}

We obtain the time of the merger from the time evolution of the distance between the centers of mass. The time of the merger is defined here as the time at which the separation is 0.5\% of its initial value.\\
Before fixing the outermost edge of our galaxies at the respective tidal radii, we investigated the dependence of the time of the merger $T_{m}$ on the cutoff radius $R_h$ of the galactic halos. In the top panel of Fig.\,\ref{time} we show that as expected, $T_{m}$  is strongly dependent on the galaxy extension only for low values of $R_h$. For $R_h>80R_d$ , the timing of the process is no longer sensitive to this parameter: from this value on, the two halos cover the entire distance between the two galaxies. \\
We also found that when $V_{t}$ increases, $T_m$ rapidly increases, as is shown in the lower panel of Fig.\,\ref{time} in the case of no IGM. It is interesting to note that there seems to be a quite accurate relation between $T_m$ and $V_{t}$: our best fit is $T_m\propto {V_{t}}^5$.\\

\begin{figure}[h]
\centering
\includegraphics[width=3.5in]{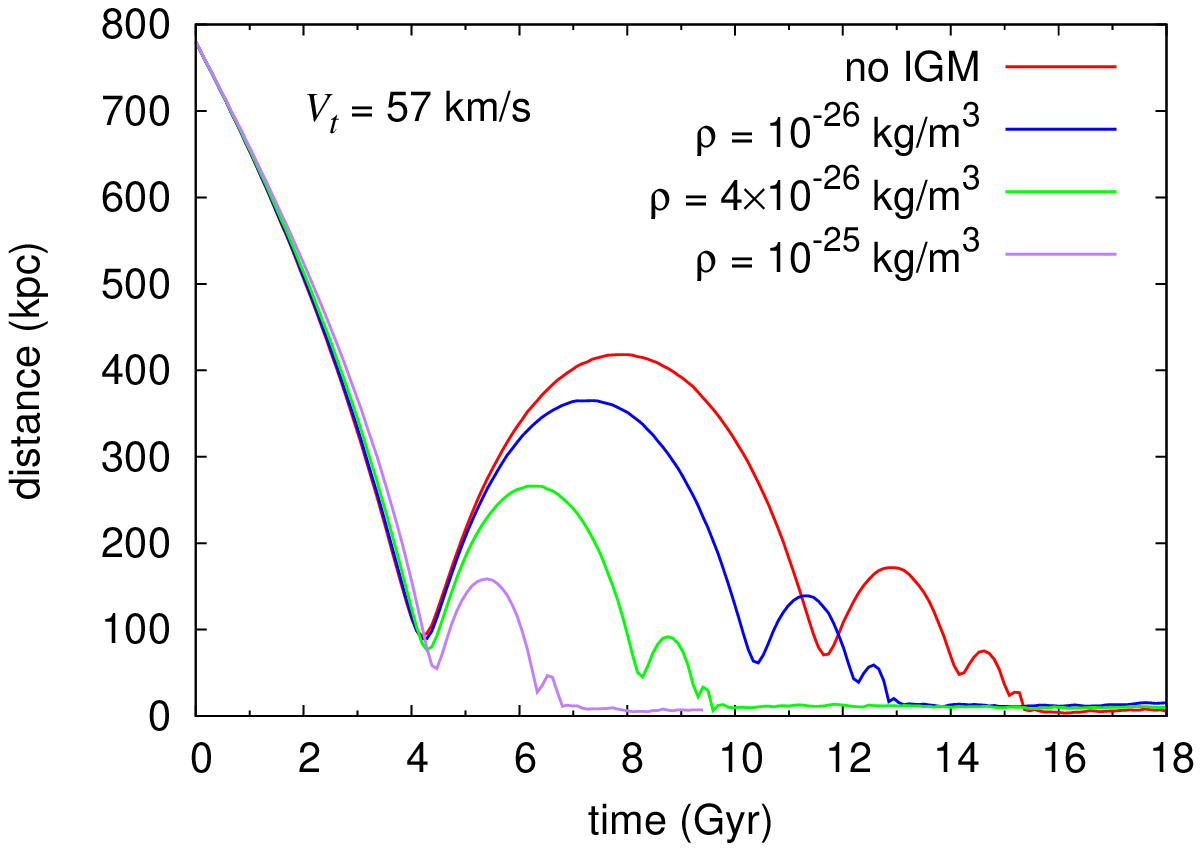}
\includegraphics[width=3.5in]{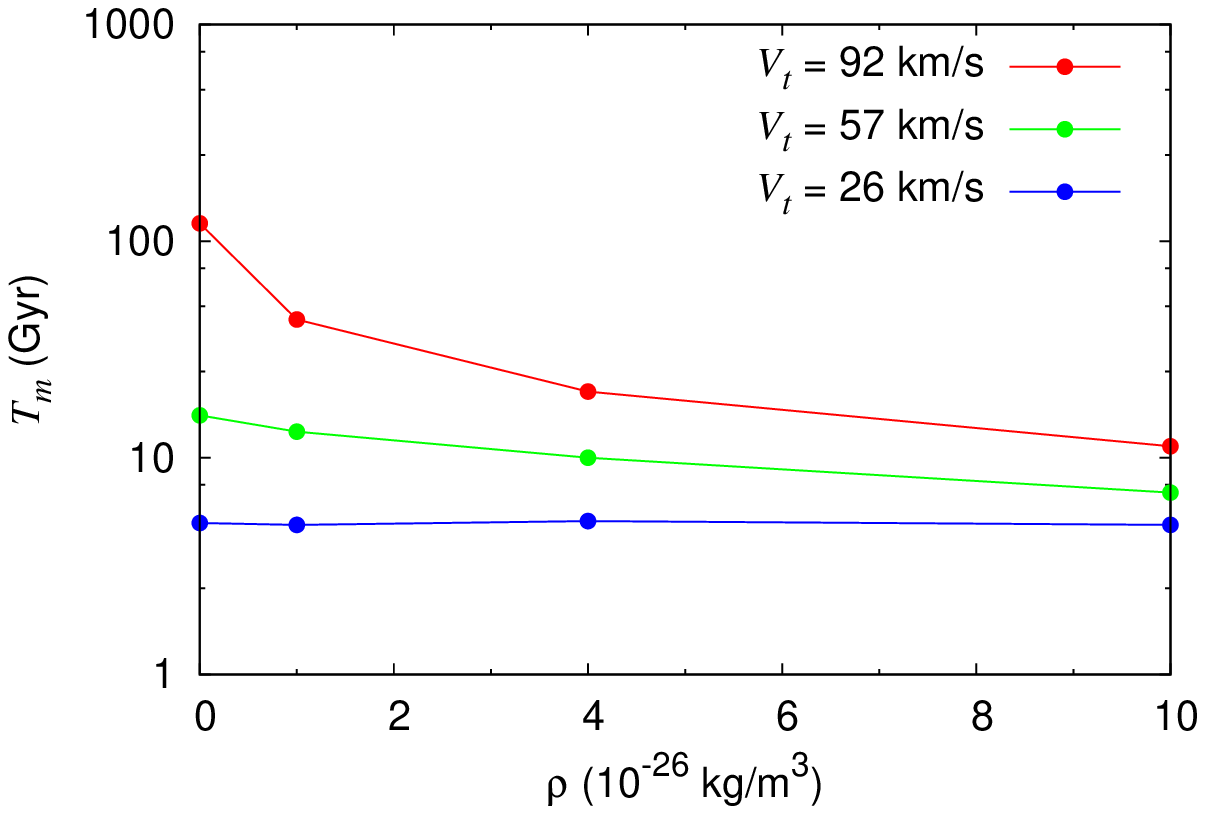}
\caption{Top: Separation between the two galaxies as a function of time for three different values of IGM density, in the case of $V_{t}=57$ km/s. Bottom: Time of the merger as a function of the IGM density for three values of $V_{t}$.}
\label{fig5}
\end{figure}

Repeating the simulation for different values of the IGM density, we obtained that the presence of a diffuse medium speeds the galaxy interaction up, especially in the case of large $V_{t}$. In Fig.\,\ref{fig5} we plot the evolution of the distance between the two centers of mass for four different values of $\rho$ in the case of $V_{t}=57$ km/s (top panel) and the dependence of the time of the merger on the IGM density for three different values of $V_{t}$ (bottom panel).\\
Even though the time of the merger can significantly change when $V_{t}$ or $\rho$ are varied, we note that the time of the first approach is almost constrained in the interval 4$-$5 Gyr. This means that the first part of the galaxy motion is nearly Keplerian because the orbital energy dissipation due to the friction exerted by the IGM is still not very efficient. The time of the first approach is very close to that obtained in the case of a pure radial fall of two point masses starting at a distance of 780 kpc with a relative radial velocity of -115.7 km/s.  After the first encounter, the IGM density instead plays a relevant role in the time for the completion of the merger. This is mainly due to the enhanced speed of the two galaxies at the pericenter. \\
Among the ensemble of the performed simulations, we refer to the simulation with $V_{t}=57$ km/s and $\rho=4.0\times10^{-26}\,\mathrm{kg/m^3}$ as a fiducial case. According to our analysis, the Milky Way and Andromeda will reach their first approach in 4.3 Gyr and will merge in 10.0 Gyr. The time for the first pericenter is close to the $\sim4.5$ Gyr found by \citet{VdM19}, but they did not report any value for the time of the final merger. \citet{Coxloeb08} obtained the first approach at $\sim2.8$ Gyr and the merger at $\sim5.4$ Gyr. However, we have to consider that they started the simulations of the MW-M31 interaction $5$ Gyr in the past and reached a current transverse velocity that is very different to the recently measured velocity. Moreover, they used an IGM density that is slightly greater than we considered in our fiducial model.

\subsection{SMBH merger} \label{smbh_merge}
\begin{figure}[h]
\centering
\includegraphics[width=3.5in]{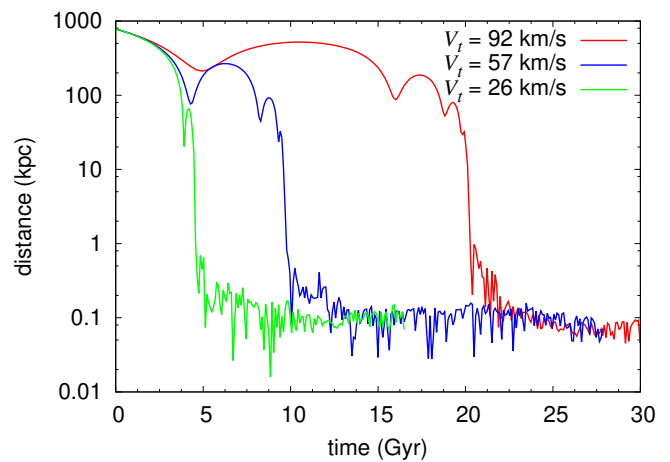}
\caption{Distance between the two SMBHs as a function of time for three different initial tangential velocities.}
\label{BHvel}
\end{figure}

\begin{figure}[h]
\centering
\includegraphics[width=3.5in]{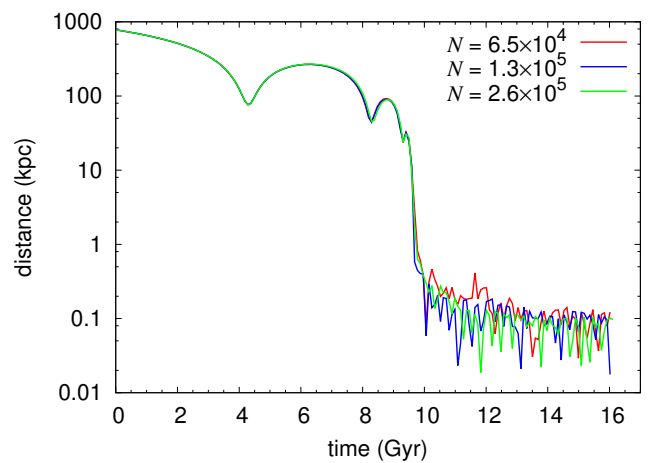}
\caption{Distance between the two SMBHs as a function of time for three different numerical resolutions of the simulation.}
\label{BHnumber}
\end{figure}

We found that the distance between the two SMBHs located at the galactic centers evolves in time, as was previously shown for the two centers of mass. The only difference is that after the galaxy merger, the SMBH binary stalls at the same fixed distance, independently of their initial velocity, as is shown in the Fig.\,\ref{BHvel}. This confirms the idea that the orbital decay of the binary in the first phase simply follows the dynamics of the two stellar systems, but it later depends on the gravitational encounters between the binary and the stars orbiting close to the galactic center. When the volume around the binary is depleted and the binary orbit contains a total mass equal to or lower than the combined mass of the two SMBHs, the binary stalls.  Therefore, as expected, this interesting behavior is a function of the number of particles that is involved in the simulation: the greater the number of particles, the smaller the stalling distance. Nevertheless, we note that as the numerical resolution increases at $N>5.0\times10^4$, the stalling distance reaches an asymptotic value of $\sim100$ pc, that is, about twice the softening parameter of the two SMBHs. Fig.\,\ref{BHnumber} shows this behavior: for three different values of $N>5.0\times10^4$, the stalling distance does not change. This might be the signal that we have reached the lower limit of the stalling distance that is allowed by our computational power. The density of stars around the binary rapidly drops to zero because of the low numerical resolution, and this makes the energy loss by dynamical friction inefficient. However, because the main purpose of this first simulation is to reproduce the galaxy interaction, we cannot expect to simultaneously correctly resolve the dynamics at small scales. The stall shown in Fig.\,\ref{BHnumber} is therefore a direct effect of the low spatial resolution and an indirect effect of the sampling.

\subsubsection{Orbital decay of the SMBH binary} \label{decay}
\begin{figure}[h]
\centering
\includegraphics[width=3.5in]{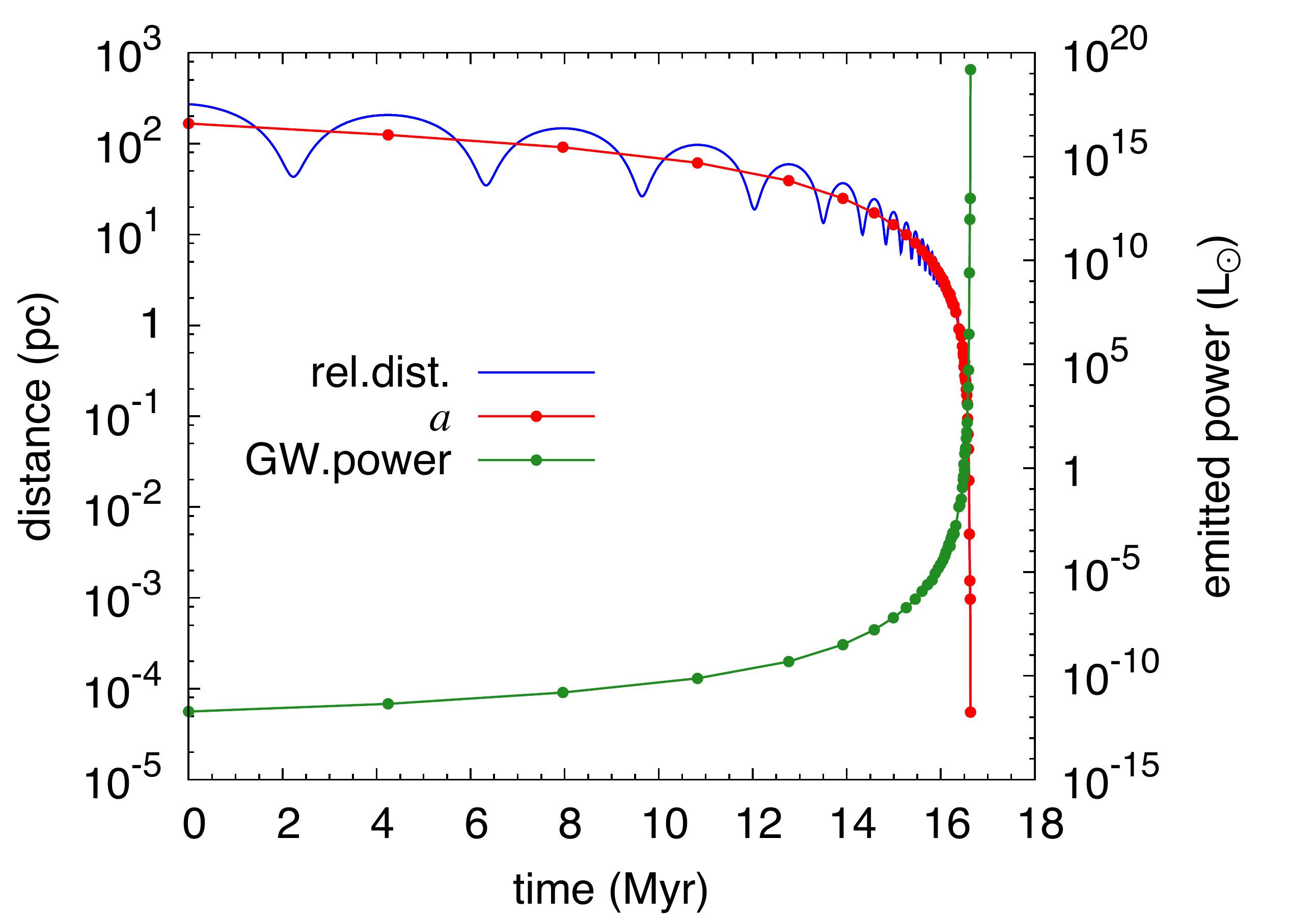}
\caption{Evolution of the relative distance between the two SMBHs (blue line) and the semimajor axis $a$ of their orbit (red line), together with the power emitted in the form of GWs (green line). The time is counted starting from the galaxy merger.}
\label{bhdecay}
\end{figure}

To follow the orbital decay of the SMBH binary, we used another code that improves the treatment of dynamical friction for the second part of our study. We chose the simulation with the largest number of particles ($N=2.6\times10^5$) and the fiducial set of parameters ($V_t=57$ km/s, $\rho=4.0\times10^{-26}\,\mathrm{kg/m^3}$) and calculated the density profile of Milkomeda at time $t=10.2$ Gyr, soon after the merger, when the system has stabilized and the SMBH binary has formed. This last point occurs when the two objects approximately reach the so-called influence radius, defined in \citet{merritt07} as the radius of a sphere around the two SMBHs that contains twice the sum of their masses. In our case, this corresponds to $r_h\approx156.5$ pc.\\
We used a Dehnen power law \citep{dehnen93}, with $\gamma=0.8$, scale length $50$ kpc, and total mass $9.3\times10^{13}\,\mathrm{M_{\odot}}$, to fit the innermost region of the density profile of Milkomeda and reproduce the galactic center as an analytic external potential in the \arwv code. The velocity dispersion in the innermost 500 parsecs ($\sigma=203.5$ km/s) was obtained from the outputs of \higpus code.\\
The values of the masses of the two SMBHs were set to the proper values $M_{MW}^{\bullet}=4.31\times10^6\,\mathrm{M_{\odot}}$, and $M_{M31}^{\bullet}=1.4\times10^8\,\mathrm{M_{\odot}}$ \citep{gillessen09,bender05}, keeping as initial conditions those coming from the last computed orbit of the SMBH pair.  
The orbital integration with the \arwv code was performed in a reference frame in which the x-y plane coincides with the initial plane of the motion.\\
After loosing orbital energy owing to the interaction with the environment, the binary becomes hard when the semimajor axis reaches the value defined in \citet{merritt07}, 
\begin{equation}
\centering
    a_h=\frac{q}{(1+q)^2}\frac{r_h}{4},
\end{equation}
where $q=M_{MW}^\bullet/M_{M31}^\bullet=0.03$ is the SMBH mass ratio. In our model, $r_h\approx156.6$ pc, and therefore $a_h\approx1.1$ pc. Fig.\,\ref{bhdecay} shows the distance between the two SMBHs and the semimajor axis of their orbit, together with the emitted GW power, as a function of time. In absolute values, the slopes of the three lines start to increase rapidly when the separation approximately reaches $a_h$. \\
Through the comparison with the case in which the calculation does not take the PN terms into account, we can infer that most of the orbital decay is due to the dynamical friction. We determined that the only interaction with the background stars can carry the binary at a distance of a few times the characteristic Schwarzschild radius of the binary, which in our case is $R_s=1.38\times10^{-5}$ pc. In the absence of PN terms, the binary would start to slowly shrink over a time of tens of million of years after this.
The emission of gravitational waves rapidly reduces the binary orbital energy and speeds up the decay: in a few thousand years, the distance plummets from few times $R_s$ to zero. According to our results, the merger between the two SMBHs occurs 16.6 Myr after the formation of Milkomeda, in the same range of timescales as  was found by \citet{Khan16} for similar SMBH mergers.

\begin{figure}[h]
\centering
\includegraphics[width=3.5in]{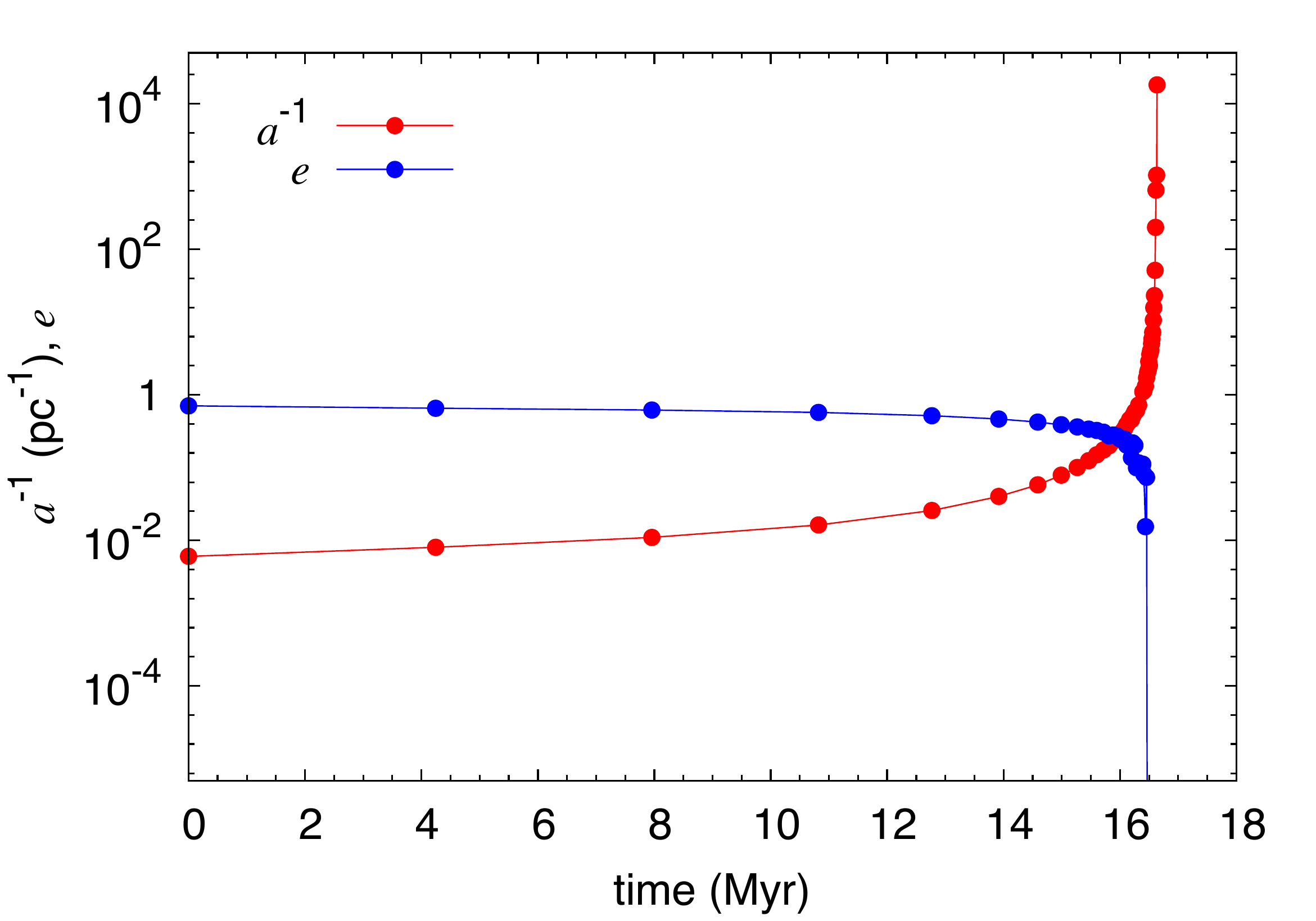}
\caption{Inverse of the semimajor axis of the orbit of the SMBHs (in $\mathrm{pc^{-1}}$) and its eccentricity as a function of time. The ordinate scale is the same for $a$ and $e$.}
\label{axis}
\end{figure}

The blue and red curves in Fig.\,\ref{axis} show the evolution of the eccentricity $e$ and the inverse semi-major axis $1/a,$ respectively. The eccentricity starts at a value of 0.7 and drops to zero, as the binary shrinks and circularizes.\\
As expected, when the binary becomes hard, the hardening rate, defined by \citet{merritt07} as
\begin{equation}
\centering
    s=\frac{d}{dt}\left(\frac{1}{a}\right),
\end{equation} 
    suddenly increases from $\sim4\,\mathrm{pc^{-1}Myr^{-1}}$ to $\sim4\times10^4\,\mathrm{pc^{-1}Myr^{-1}}$ in the last phases before the merger.\\ However, because the external environment in \arwv is not modeled with an ensemble of particles, we cannot reproduce the orbital energy loss due to the stochastic gravitational encounters of the binary with the field stars.
Our estimation of $s$, and of $|de/dt|$ as well, in the phases soon after the binary becomes hard is therefore underestimated. We can obtain an estimate of the hardening rate due to the energy exchange during the encounters, in the assumption of a background with a fixed and uniform density $\rho$ and uniform velocity $v$, through the approximated formula \citep{Quinlan_1996,Gualandris_2016} 
\begin{equation}
s=\frac{G\rho}{v}H,
\end{equation}
where $H$ is a dimensionless hardening coefficient with a nearly constant value of $\sim16$ for hard binaries. In our case, this is $s\approx0.21\,\mathrm{pc^{-1}Myr^{-1}}$, in accordance with those obtained in similar simulations by \citet{Khan2018}.

\subsubsection{Gravitational wave emission} \label{GW}

The power emitted in the form of GWs as a function of time, shown with the green line in Fig.\,\ref{bhdecay}, was obtained as the energy loss rate, averaged over one orbital period, according to the formula (16) of \citet{peters63},
\begin{equation}
\label{peters}
\centering
    \langle P \rangle=\frac{32}{5}\frac{G^4m_1^2m_2^2\left(m_1+m_2\right)}{c^5a^5\left(1-e^2\right)^{7/2}}\left(1+\frac{73}{24}e^2+\frac{37}{96}e^4\right).
\end{equation}

\begin{figure}[h]
\centering
\includegraphics[width=3.5in]{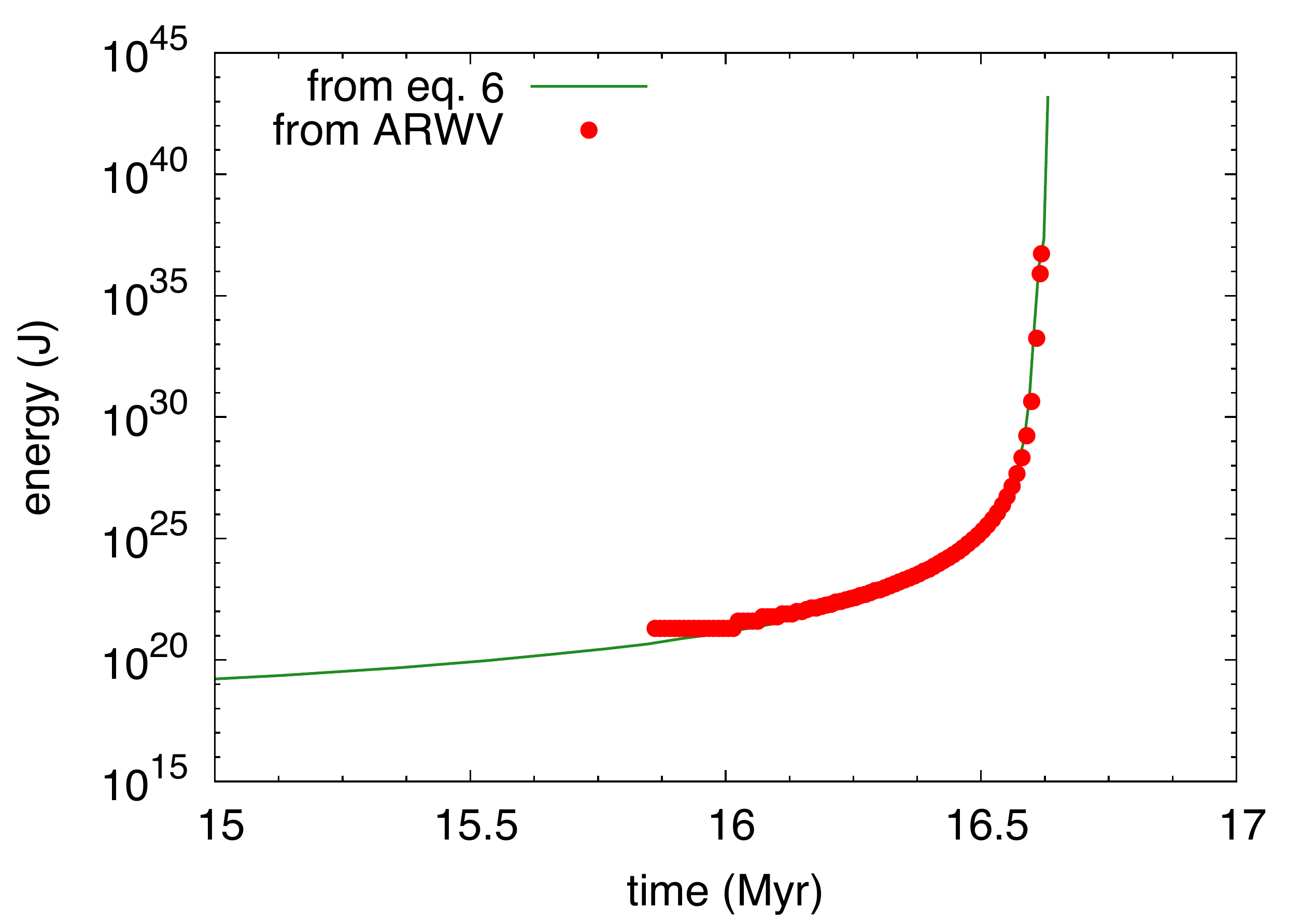}
\caption{Energy radiated during the last phases of the SMBH merger. The green line shows the mean emitted energy, obtained from the integration of Eq. \eqref{peters}, and red points show the emitted energy as output of the \arwv code.}
\label{fig10}
\end{figure}

\begin{figure}[h]
\centering
\includegraphics[width=3.5in]{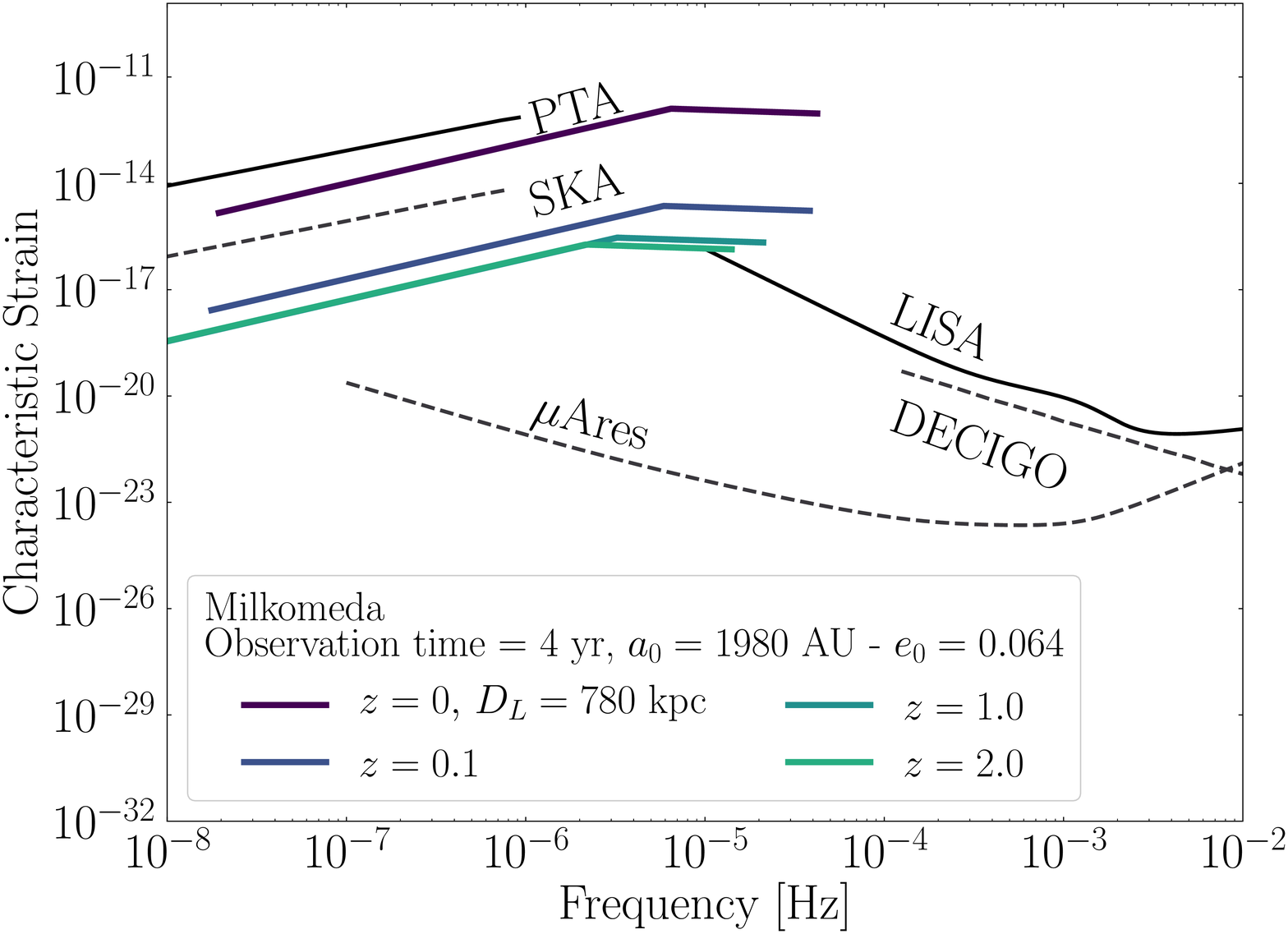}
\caption{Characteristic strain of a GW signal emitted by a similar SMBH merging binary as a function of frequency for four different redshifts. The GW signal is computed starting from a semimajor axis of $a_0=1980$ AU and an eccentricity of $e_0=0.064$, corresponding to a time of $t=4$ yr before the merger. The sensibility curves of the main GW detectors are also shown.}
\label{strain}
\end{figure}

The emitted power progressively increases  when  the  semimajor  axis  drops  below $a_h$ and reaches a maximum of about $10^{19}\,\mathrm{L_{\odot}}$ just before the merger.\\
Through a numerical integration of the Peters formula, we obtained the amount of energy that is radiated away during the process. In Fig.\,\ref{fig10} we compare the results of this integration with the energy loss calculated by the \arwv code through the PN approximation. The two curves agree well, with a fractionary logarithmic variation below $\sim13\%$. The amount of energy emitted in the last phases of process is on the order of $10^{43}$ J.\\ 
We repeated the simulation with the \arwv code for two extreme configurations of the SMBH spin vectors: one for parallel spins, and the other for antiparallel spins. As expected, the time of the merger is not affected by the spin orientation, but we found that the final merger remnant gains a different recoil velocity. In the case of parallel spins, the recoil velocity of the remnant is $v_{r}=2.2$ km/s, while for antiparallel spins, it is $v_r=24.8$ km/s. The magnitude of the recoil velocity in both cases is small, mainly because of the low SMBH mass ratio \citep{hea18,cha20}. The final SMBH is thus unable to escape from the center of the galaxy because the central escape velocity for our model is $v_e=3.7\times10^3$ km/s. Therefore, the giant elliptical galaxy Milkomeda will continue to host an SMBH in its center, as obtained also by \citet{arcasedda19b}.\\
We also investigated the possibility of observing a GW signal from a merger between similar SMBHs in the near Universe. The frequency-characteristic strain evolution is shown in Fig.\,\ref{strain} and overlaps the sensitivity curve of different GW detectors, such as the Pulsar Timing Array (PTA, \citet{Hobbs10}), the Square Kilometer Array (SKA, \citet{Johnston07}), the Laser Interferometer Space Antenna (LISA, \citet{Amaro17}), the Deci-Hertz Interferometer Gravitational wave Observatory (DECIGO, \citet{Kawamura11}), and the $\mu$Ares microhertz detector \citep{Sesana19}. A comparison between our modeled signal and the detector sensitivities shows that mergers similar to the one we expect to witness in Milkomeda can be bright sources in ground-based detectors such as the PTA, or in the next decade, the SKA, provided that they take place roughly within 1 Mpc. However, farther away in space, it becomes clear that the only possibility of observing this type of SMBH mergers is using space-borne detectors. Mergers that occur up to redshift $z\lesssim 2$ might be observable by LISA, if with an allegedly low signal-to-noise ratio, but they might shine bright to a micro-hertz observatory such as the $\mu$Ares detector concept design.

\section{Discussion and conclusions}
We studied the future evolution of the system composed of the Milky Way and M31 (Andromeda galaxy) in relation with the gravitational effects due to the intergalactic background. This can shed light not only on the forthcoming dynamics of our specific galaxy system, but also in general on the correlation between the galaxy interaction timing and the environmental properties, the galactic structure, and the initial conditions.  We used the high-precision $N$-body \higpus code, properly modified to account for the friction exerted on the bodies in the galaxies by the diffuse background, to calculate the evolution of the galactic hosts and their central SMBHs until the merger of the two galaxies. At this stage, we used the galaxy density, the velocity distribution, and the SMBH orbital parameters for a further dynamical simulation. With the \arwv code in its most recent version, which considers PN treatment and the BH spins, we reproduced the orbital decay of the SMBH binary in the post-merger galactic background.
The aim was not simply to estimate the likelihood and future time of an eventual merger, but also to determine the fate of the two massive black holes hosted at the two galactic centers, whose estimated mass is $4.31\times 10^6$ M$_\odot$ in our Galaxy and $1.4\times 10^8$ M$_\odot$ in M31, giving a mass ratio $0.03$.
We summarize our results below.

\begin{enumerate}
    \item The time evolution of the MW and M31 orbits is such that the first close approach of the two galaxies will occur in $4-5$ Gyr, with a weak dependence on the characteristics assumed for the background density, the dimension of the halos, and the initial velocity. 
    \item The time for the completion of the two galaxy merger increases significantly with the relative velocity transverse component, which is an ill-determined observational quantity. According to the most recent estimates, however we can conclude that the MW and M31 will merge in $\sim10$ Gyr.
    \item The dimensions of the galactic halos play an important role in the merger time: larger halos cause a significant orbital energy dissipation and accelerate the decay, at least until the distance between the two galaxies is on the same order as the size of the halos.
    \item As expected, due to the collisionless nature of the encounter and merger, the overall post-merger density profile is not very different from a mere mass average of the two profiles of the MW and M31.
    \item After the merger, the two SMBHs were left orbiting on a  mutual orbit of eccentricity $e\sim 0.7$ and semimajor axis $a\sim 160$ pc, which stalled because the resolution of the $N$-body simulation is insufficient.
    \item The following fate of the SMBH pair was followed by a PN simulation that showed how efficiently (in less than $17$ Myr) dynamical friction braking leads the two SMBHs to the so-called hard binary phase, when subsequent orbital decay is given by energy dissipation by GW emission down to the final merger and recoil kick.
    \item When we also considered antiparallel spins for the SMBHs, the recoil kick velocity was below $25$ km/s (two orders of magnitude lower than the central escape velocity), which leaves the BH remnant confined in the inner potential well of the galaxy.
    \item Types of SMBH orbital decays similar to those studied here show a very high power of GW emission because of the high masses of the BH involved. This high power is shifted toward very low frequency. This GW emission would in principle be observable only with future GW ground-based detectors such as the PTA and the SKA or with space interferometers such as LISA, but the redshift range for the detection should be $1\leq z\leq 2$.
\end{enumerate}

On the basis of our results, we are now able to determine the most feasible scenario of the future of our own Galaxy and its central SMBH. Our new estimate of the time required for the completion of the MW-M31 merger means that the life of the Local Group is slightly longer than previously believed. A final result is that unfortunately, the Sun will not live long enough to witness the formation of Milkomeda and will therefore not be part of the new galaxy.

\bibliographystyle{aa} 
\bibliography{references} 

\end{document}